\documentclass[preprints,review,accept,moreauthors]{Definitions/mdpi} 

\firstpage{1} 
\pubvolume{1}
\issuenum{1}
\articlenumber{0}
\pubyear{2026}
\copyrightyear{2026}
\datereceived{ } 
\daterevised{ } 
\dateaccepted{ } 
\datepublished{ } 
\longauthorlist{yes} 

\usepackage{graphicx}
\Title{Status and Progress of Hamiltonian Effective Field Theory in Light Hadron Spectroscopy}

\Author{Qi-Liang Xie $^{1,2,3}$, Qiu-Rui Wu $^{1,2,3,*}$, Zi-Jun Li $^{1,2,3,*}$, Fang-Chao Han $^{1,2,3,*}$, \mbox{Yu Zhuge $^{1,2,3,*}$} and Zhan-Wei Liu $^{1,2,3,*}$}

\AuthorNames{Qi-Liang Xie, Qiu-Rui Wu, Zi-Jun Li, Fang-Chao Han, Yu Zhuge and Zhan-Wei Liu}

\address{%
$^{1}$ \quad School of Physical Science and Technology, Lanzhou University, Lanzhou 730000, China\\
$^{2}$ \quad Research Center for Hadron and CSR Physics, Lanzhou University and Institute of Modern Physics of CAS, Lanzhou 730000, China\\
$^{3}$ \quad Lanzhou Center for Theoretical Physics, MoE Frontiers Science Center for Rare Isotopes,
Key Laboratory of Quantum Theory and Applications of MoE,
Key Laboratory of Theoretical Physics of Gansu Province,
Gansu Provincial Research Center for Basic Disciplines of Quantum Physics, Lanzhou University, Lanzhou 730000, China}

\corres{Correspondence: wuqr2025@lzu.edu.cn; lzijun2024@lzu.edu.cn; hanfch2023@lzu.edu.cn; zhugy2025@lzu.edu.cn; liuzhanwei@lzu.edu.cn}

\abstract{Hadron resonances are intrinsically dynamical states whose observable properties are governed by the interplay between quark-model configurations and hadronic channels. Hamiltonian effective field theory provides a practical framework for connecting hadron dynamics with experimental observables and lattice QCD results. In this review, we outline the infinite- and finite-volume formulations of Hamiltonian effective field theory and survey its applications to light-hadron spectroscopy and resonance structure. By constraining the same Hamiltonian framework with experimental scattering data and lattice QCD spectra, Hamiltonian effective field theory can help distinguish among scenarios that describe the available scattering data comparably well but imply different roles for bare-state and multihadron configurations. We also review recent extensions of the Hamiltonian framework developed to address more complicated dynamical settings.}

\begin{document}

\section{Introduction}
Light-baryon resonances have been studied experimentally for decades. The $N^*(1440)$, commonly known as the Roper resonance, was identified in early partial-wave analyses of pion-nucleon scattering in 1964 \cite{Adelman:1964jmp, Bareyre:1964pcg, Roper:1964zza}. The quark model provides a useful phenomenological description of baryon spectra. It accounts for many observed states; however, with the ever-increasing number of observed resonances, certain phenomena and exotic hadronic states have emerged that cannot be accommodated within the naive quark model. Notable examples include the mass inversion problem of the $N^*(1440)$ and $N^*(1535)$ states, as well as the fact that the masses of the $\Xi(1620)$ and $\Xi(1690)$ are significantly lower than the quark-model prediction of about $1800\,\text{MeV}$ \cite{Belle:2018lws}. In addition, exotic multi-quark states have also been reported \cite{LHCb:2015yax, LHCb:2019kea, LHCb:2021vvq}. Elucidating these systems provides an essential avenue toward understanding the internal structure of hadrons. Meanwhile, these resonance states also provide an important platform for investigating the non-perturbative regime of quantum chromodynamics (QCD).

Taking the nucleon as an example, excited nucleon resonances are commonly studied through pion-nucleon scattering and electromagnetic production reactions \cite{Doring:2025wms}. These resonances are strongly coupled to two-particle meson-baryon channels \cite{ParticleDataGroup:2026aaa}. They can be viewed either as dynamically generated by the coupled meson-baryon interactions, or as states that also contain an intrinsic bare-core contribution \cite{Ramalho:2023hqd}. In partial-wave analyses, contributions from both types are usually taken into account simultaneously \cite{Matsuyama:2006rp, Briscoe:2023gmb, Drechsel:2007if, Ronchen:2014cna}. In this approach, a phenomenological model is constructed and fitted to experimental scattering data, so as to extract the pole positions of the resonances. This framework has been widely applied to the study of low-lying baryon excitations.

Currently, lattice QCD provides finite-volume energy levels and matrix elements from first principles, making it a crucial tool for studying hadronic resonances \cite{Edwards:2011jj, Edwards:2012fx, Kiratidis:2015vpa, Mahbub:2013ala, Alexandrou:2014mka, Bulava:2022vpq, Hockley:2023yzn, Menadue:2011pd, Hall:2014uca, BaryonScatteringBaSc:2023ori}. Scattering parameters and resonance poles are then inferred using finite-volume relations and analytic continuation. However, owing to limitations in computational resources, it is necessary to extrapolate the lattice QCD results to the continuum limit and the physical quark-mass region. The Lüscher method \cite{Luscher:1986pf} establishes a rigorous relation between the discrete energy spectrum in a finite volume and the infinite-volume scattering amplitude, thereby enabling the extraction of the pole positions of resonances from the lattice energy levels. At present, this framework is being extended to more complex scenarios, including systems involving particles with non-zero spin, coupled-channel interactions, and three-body final states \cite{Briceno:2017max}.

Hamiltonian effective field theory (HEFT) is a useful approach for connecting lattice QCD results in the unphysical regime with experimental data in the physical region \cite{Hall:2013qba, Wu:2014vma, Liu:2015ktc, Liu:2016uzk, Liu:2016wxq, Wu:2016ixr, Wu:2017qve, Li:2019qvh}. HEFT describes coupled-channel dynamics using effective field theory and phenomenological input and is consistent with the Lüscher relation in the two-body regime~\cite{Luscher:1986pf}. By constructing the Hamiltonian of the system and solving the coupled-channel equations, scattering observables such as inelasticity, phase shifts, and cross sections can be obtained. Applying the Hamiltonian to a finite volume yields the energy spectrum, which can be compared with lattice QCD spectra, thereby helping us understand the structure of baryon resonances.

HEFT has been widely applied in the study of hadronic resonances. Initially, it was used to investigate the $\Delta(1232)$, revealing that its eigenstate is a mixture of the bare state and the $N\pi$ multi-particle state \cite{Hall:2013qba}. Subsequently, HEFT was extended from the single-channel framework to more complex two-channel and even multichannel scattering systems, thereby providing a practical parametrization for analyzing limited coupled-channel data~\cite{Wu:2014vma, Yu:2026qlt}. It has also been generalized to handle cases with partial-wave mixing \cite{Li:2019qvh, Li:2021mob}. HEFT has been employed to analyze the structures of the $N^*(1535)$ and $N^*(1650)$ \cite{Liu:2015ktc, Abell:2023nex}, and offers a dynamical explanation of the mass inversion of the $N^*(1440)$ Roper resonance \cite{Liu:2016uzk, Wu:2017qve} as well as that of the $\Delta(1600)$ \cite{Hockley:2024ipz}. Finite-volume analyses support a predominantly $\bar{K}N$ molecular interpretation of $\Lambda(1405)$ \cite{Hall:2014uca, Liu:2016wxq}, whereas the $\Lambda(1670)$ contains a significant three-quark component within the model \cite{Liu:2023xvy}. The analysis of the $\Omega(2012)$ favors the assignment $J^P=\frac{3}{2}^-$ and describes it as a coupled system containing both a three-quark core and meson-baryon components \cite{Han:2025gkp}. In the negative-parity $\Sigma$ sector, the available lattice spectra favor a scenario containing a bare state and two resonance poles near $1.7~\mathrm{GeV}$, which can be associated with the experimentally observed $\Sigma(1620)$ and $\Sigma(1750)$ \cite{Ma:2025wbz}. These results are supported by comparisons with lattice QCD spectra, greatly advancing our understanding of the internal structures of resonances.

This paper is organized as follows. In Sec.~\ref{seca}, we describe the theoretical framework of HEFT, covering both the infinite-volume and finite-volume formulations. In Sec.~\ref{secb}, we present applications of HEFT to the study of resonance structures, with examples including the $\rho(770)$, the $N^*$, $\Delta$, $\Lambda$, $\Sigma$ families, and the $\Omega(2012)$. In Sec.~\ref{secc}, we review the application of HEFT to more complex systems, such as N-body systems and photoproduction/electroproduction processes. Finally, in Sec.~\ref{secd}, we give a brief summary.

\section{Hamiltonian Effective Field Theory}
\label{seca}
\subsection{Infinite-volume HEFT}

HEFT allows for the construction of a Hamiltonian in infinite volume, which can be constrained by experimental data such as cross sections, inelasticities, and phase shifts.  

\subsubsection{Hamiltonian}

The Hamiltonian can be divided into two parts,
\begin{equation}
\label{Hamiltonian}
H=H_0+H_I \, ,
\end{equation}
where $H_0$ represents the free part and $H_I$ is the interacting part. The non-interacting Hamiltonian can be written as
\begin{eqnarray}
\label{H0Hamiltonian}
H_0&=&\sum_{B_0}|B_0\rangle\, m_B^0\,\langle B_0| +\sum_{\alpha}\int d^3 k\left|\alpha\left(k\right)\right\rangle\left[\omega_{\alpha_{M}}\left(k\right)+\omega_{\alpha_B}\left(k\right)\right]\left\langle\alpha\left(k\right)\right|.\quad 
\end{eqnarray}
Here, $m_B^0$ is the mass of the bare baryon $B_0$, $\alpha_{M}$($\alpha_{B}$) labels the meson (baryon) in channel $\alpha$, and $\omega_{\alpha_{M\left(B\right)}}\left(k\right)=\sqrt{m^2_{\alpha_{M\left(B\right)}}+k^2}$ refers to the meson (baryon) energy.

There are two types of vertex interactions in the interacting Hamiltonian 
\begin{equation}
\label{HamiltonianI}
H_I=g+v \, ,
\end{equation}
where $g$ denotes the coupling of the bare state $B_0$ to the meson-baryon channel $\alpha$
\begin{eqnarray}
\label{bareHamiltonian}
g=\sum_{B_0,\alpha}\int d^3 k\left\{\left|B_0\right\rangle G^{\dagger}_{B_0,\alpha}\left(k\right)\left\langle \alpha\left(k\right)\right|+h.c.\right\} \, ,
\end{eqnarray}
$v$ represents the direct two-to-two particle interaction
\begin{eqnarray}
v=\sum_{\alpha,\beta}\int d^3 kd^3 k^\prime\left|\alpha(k)\right\rangle V_{\alpha,\beta}\left(k,k^\prime\right)\left\langle\beta(k^\prime)\right| \, .
\end{eqnarray}

A commonly used form for the S-wave one-to-two interaction is
\begin{eqnarray}
\label{bareintaction}
G_{B_0,\alpha}(k)=\frac{\sqrt{3}\,g_{B_0, \alpha}}{2\pi f}\sqrt{\omega_{\alpha_M}(k)}\,u(k) \, ,
\end{eqnarray}
and the non-separable form of the S-wave two-to-two potentials can be derived from the Weinberg-Tomozawa interaction~\cite{PhysRevD.31.1033}:
\begin{eqnarray}
V_{\alpha,\beta}\left(k,k^\prime\right)=v_{\alpha,\beta}\frac{\left[\omega_{\alpha_M}\left(k\right)+\omega_{\beta_M}\left(k^\prime\right)\right]u\left(k\right)u\left(k^\prime\right)}{8\pi^2f^2\sqrt{2\omega_{\alpha_M}\left(k\right)}
	\sqrt{2\omega_{\beta_M}\left(k^\prime\right)}} \, .
\label{interactionp}
\end{eqnarray}
Separable potentials are also used~\cite{Wu:2017qve}:
\begin{eqnarray}
V_{\alpha,\beta}(k, k') = \frac{v_{\alpha,\beta}}{4\pi^2} \left(\frac{k}{f}\right)^{l_\alpha} \left(\frac{k'}{f}\right)^{l_\beta} \frac{u_\alpha(k)}{\omega_{\alpha_M}(k)} \frac{u_\beta(k')}{\omega_{\beta_M}(k')} \, .
\end{eqnarray}
Here, the dipole form factor $u(k)=(1+k^2/\Lambda^2)^{-2}$ and the Gaussian form factor $u(k) = \exp\left(-k^2/{\Lambda^2}\right)$ are commonly used. The cutoff $\Lambda$ may depend on the channel, in which case the regulator is denoted by $u_\alpha(k)=u(k;\Lambda_\alpha)$. $f$ denotes the pion decay constant and $l_{\alpha}$ the orbital angular momentum in channel $\alpha$. The parameters $g_{B_0, \alpha}$ and $v_{\alpha,\beta}$ denote the coupling constants of the corresponding interactions.

\subsubsection{T-matrix}

One can obtain the T-matrices for two-particle scattering by solving the three-dimensional reduction of the coupled-channel Bethe-Salpeter equation
\begin{eqnarray}
T_{\alpha,\beta}(k,k^\prime;E)&=&\widetilde{V}_{\alpha,\beta}(k,k^\prime;E)+\sum_\gamma
\int q^2dq\frac{\widetilde{V}_{\alpha,\gamma}(k,q;E)\, T_{\gamma,\beta}(q,k^\prime;E)}{E-\omega_{\gamma_1}(q)-\omega_{\gamma_2}(q)+i\epsilon},
\end{eqnarray}
where the coupled-channel potential can be written in terms of the interacting Hamiltonian above as
\begin{eqnarray}
\widetilde{V}_{\alpha,\beta}(k,k^\prime ;E)&=&\sum_{B_0}\frac{G^{\dagger}_{B_0,\alpha}(k)\,G_{B_0,\beta}(k^\prime)}
{E-m_B^0}+V_{\alpha,\beta}(k,k^\prime) \, .
\end{eqnarray}

\subsubsection{Scattering Observables}

With the T-matrices determined from the Bethe-Salpeter equation, one can compute various scattering observables, such as the cross section describing the process from channel $\beta$ to channel $\alpha$
\begin{equation}
\sigma_{\alpha, \beta} =
\frac{4\pi^3 \, k_{\mathrm{cm}}^{\alpha} \, 
\omega_{\alpha_M}^{\mathrm{cm}} \, 
\omega_{\alpha_B}^{\mathrm{cm}} \, 
\omega_{\beta_M}^{\mathrm{cm}} \, 
\omega_{\beta_B}^{\mathrm{cm}}}
{E_{\mathrm{cm}}^2 \, k_{\mathrm{cm}}^{\beta}}
\left| T_{\alpha, \beta}
\!\left( k_{\mathrm{cm}}^{\alpha}, k_{\mathrm{cm}}^{\beta}; E_{\mathrm{cm}} \right) \right|^2. 
\end{equation}
The relation between the T-matrix and the S-matrix is defined by
\begin{equation}
S_{\alpha\beta}(E) = \delta_{\alpha\beta} - 2i\pi \sqrt{\rho_{\alpha}\rho_{\beta}} T_{\alpha\beta}(k_{\alpha}, k_{\beta}; E) ,
\end{equation}
where the phase-space factors $\rho_{\alpha}$ are 
\begin{equation}
\rho_{\alpha} = \frac{\sqrt{k_{\alpha}^2 + m_{\alpha M}^2} \sqrt{k_{\alpha}^2 + m_{\alpha B}^2}}{E} k_{\alpha},
\end{equation}
from which one can extract the phase shift $\delta_{\alpha\beta}$ and the inelasticity $\eta_{\alpha\beta}$ via
\begin{equation}
S_{\alpha\beta}(E) = \eta_{\alpha\beta}(E) \exp\!\bigl( 2 i \delta_{\alpha\beta}(E) \bigr) , 
\end{equation}
\begin{equation}
\delta_{\alpha\beta}(E) = \frac{1}{2} \arctan \left( \frac{\text{Im}\{S_{\alpha\beta}(E)\}}{\text{Re}\{S_{\alpha\beta}(E)\}} \right), \quad \eta_{\alpha\beta}(E) = |S_{\alpha\beta}(E)|.
\end{equation}
Here the usual channel observables correspond to $\alpha=\beta$, and the inverse tangent is understood with quadrant correction and continuous phase tracking.

\subsection{Finite-volume HEFT}

In order to connect our numerical results with lattice QCD simulations, we need to extend HEFT to finite volume. In a box with length $L$, the momentum of a particle becomes discretized as $k_n=2\pi\sqrt{n}/L$, with $n=n_x^2+n_y^2+n_z^2$, where $n=0,1,2,\dots$.

\subsubsection{Hamiltonian Matrix}

With $\omega_\alpha(k)=\omega_{\alpha M}(k)+\omega_{\alpha B}(k)$, the free Hamiltonian can be written as
\begin{equation}
\mathcal{H}_{0}
= \mathrm{diag}\bigl\{
m_{B}^{0}, 
\omega_{\alpha}(k_{0}), 
\omega_{\beta}(k_{0}), 
\cdots, 
\omega_{\alpha}(k_{1}), 
\cdots
\bigr\}  , 
\end{equation}
and the interacting Hamiltonian is
\begin{equation}
\scalebox{1.0}{$\mathcal{H}_{I}
=
\begin{pmatrix}
0
& \mathcal{G}^{\dagger}_{B_{0}, \alpha}(k_{0})
& \mathcal{G}^{\dagger}_{B_{0}, \beta}(k_{0})
& \cdots
& \mathcal{G}^{\dagger}_{B_{0}, \alpha}(k_{1})
& \cdots \\[6pt]

\mathcal{G}_{B_{0}, \alpha}(k_{0})
& \mathcal{V}_{\alpha, \alpha}(k_{0}, k_{0})
& \mathcal{V}_{\alpha, \beta}(k_{0}, k_{0})
& \cdots
& \mathcal{V}_{\alpha, \alpha}(k_{0}, k_{1})
& \cdots \\[6pt]

\mathcal{G}_{B_{0}, \beta}(k_{0})
& \mathcal{V}_{\beta, \alpha}(k_{0}, k_{0})
& \mathcal{V}_{\beta, \beta}(k_{0}, k_{0})
& \cdots
& \mathcal{V}_{\beta, \alpha}(k_{0}, k_{1})
& \cdots \\[6pt]

\vdots & \vdots & \vdots & \ddots & \vdots & \ddots \\[6pt]

\mathcal{G}_{B_{0}, \alpha}(k_{1})
& \mathcal{V}_{\alpha, \alpha}(k_{1}, k_{0})
& \mathcal{V}_{\alpha, \beta}(k_{1}, k_{0})
& \cdots
& \mathcal{V}_{\alpha, \alpha}(k_{1}, k_{1})
& \cdots \\[6pt]

\vdots & \vdots & \vdots & \ddots & \vdots & \ddots
\end{pmatrix}$},
\end{equation}
where
\begin{eqnarray}
\mathcal{G}_{B_0,\alpha}(k_n)&=&\sqrt{\frac{C_3\left(n\right)}{4\pi}}\left(\frac{2\pi}{L}\right)^{3/2}G_{B_0,\alpha}\left(k_n\right),\\
\mathcal{V}_{\alpha,\beta}\left(k_n,k_m\right)&=&\frac{\sqrt{C_3\left(n\right)C_{3}\left(m\right)}}{4\pi}\left(\frac{2\pi}{L}\right)^3V_{\alpha,\beta}\left(k_n,k_m\right) \, .\qquad
\end{eqnarray}
$C_{3}(n)$ denotes the number of ways $n$ can be expressed as the sum of the squares of three integers.

\subsubsection{Pion Mass Dependence}

Since lattice QCD calculations are typically performed at pion masses above the physical point, we need to extend HEFT to unphysical pion masses. The pion mass dependence for the meson and baryon is given by
\begin{align}
&m_M(m_{\pi}^2)=\sqrt{m_M^2|_{phy}+\alpha_M(m_{\pi}^2-m_{\pi}^2|_{phy})},\\
&m_B(m_{\pi}^2)=m_B|_{phy}+\alpha_B(m_{\pi}^2-m_{\pi}^2|_{phy}).
\end{align}
Here, $\alpha_M$ and $\alpha_B$ denote the meson mass-squared slope and the baryon mass slope, respectively, obtained by fitting to lattice QCD results. In this way, we extend our numerical results to unphysical points, enabling a direct comparison with lattice QCD calculations. Furthermore, we also need to extend the bare mass to unphysical points. By solving the eigenvalue equation of the Hamiltonian, we finally obtain the finite-volume eigenvalue spectrum and eigenvector components, which allow us to probe the structure of light hadrons through direct comparison with lattice QCD data and the analysis of eigenvector composition. Some illustrative examples are given below.

\section{Representative Applications of HEFT}
\label{secb}
 \subsection{$\rho(770)$}
Although the $\rho(770)$ meson is generally regarded as a conventional $q\bar q$ vector meson, its experimentally observed width of approximately $150~\mathrm{MeV}$~\cite{ParticleDataGroup:2026aaa,Protopopescu:1973sh,Hyams:1973zf} indicates a strong coupling to the $\pi\pi$ continuum. In the past decade, multiple LQCD groups, including HSC~\cite{Dudek:2012xn,Wilson:2015dqa}, PACS-CS~\cite{CS:2011vqf}, MILC~\cite{Fu:2016itp}, ETMC~\cite{ExtendedTwistedMass:2019omo}, Guo et al.~\cite{Guo:2016zos}, C. Alexandrou et al.~\cite{Alexandrou:2017mpi} and J. Bulava et al.~\cite{Andersen:2018mau} have provided energy levels for the $I=1$ P-wave sector, with different pion masses. However, there has been little work collating spectra from various collaborations. To synthesize these disparate multi-collaboration datasets, Yu et al. applied HEFT to perform a comprehensive global analysis of $\rho$-meson properties~\cite{Yu:2023xxf}. Their theoretical setup incorporates a bare $\rho$-meson basis state representing the $q\Bar{q}$ quark-model core, coupled to $\pi\pi$ and $\pi\omega$ channels.

In this work, a unified analysis indicated that both the coupling constant $g_{\rho\pi\pi}$ and the regulator cutoff $\Lambda_{\rho\pi\pi}$ associated with the $\rho\text{--}\pi\pi$ vertex exhibit only a weak dependence on $m_\pi$. This behavior suggests that the vertex remains relatively stable under chiral extrapolation. Meanwhile, the bare $\rho$-meson mass $m_\rho^B$ displays an approximately linear dependence on $m_\pi^2$, which the authors interpret as being consistent with a predominantly valence-quark core.

Using the five eigenstates obtained from the spectrum generated with the HSC (2015) data~\cite{Wilson:2015dqa} at $m_\pi=236~\mathrm{MeV}$, the authors found that the model including $\pi\pi$ and $\pi\omega$ channels gives bare-$\rho$ probabilities of approximately 62\%--75\%, $\pi\pi$ probabilities of 20\%--34\%, and $\pi\omega$ probabilities of about 4\%~\cite{Yu:2023xxf}. These model-dependent fractions support a substantial quark-model core with non-negligible meson-cloud dressing.

\subsection{{$N^*(1440)$, $N^*(1535)$ and $N^*(1650)$}}
The Roper resonance, $N^*(1440)$, was first identified in 1964 through a partial-wave analysis of $\pi N$ scattering in the $P_{11}$ channel~\cite{Adelman:1964jmp,Bareyre:1964pcg, Roper:1964zza}. The $N^*(1535)$ and $N^*(1650)$ were subsequently established as resonance states in the $S_{11}$ $\pi N$ partial wave~\cite{ParticleDataGroup:2026aaa,Matsuyama:2006rp, Briscoe:2023gmb, Drechsel:2007if, Ronchen:2014cna}. A longstanding structural puzzle persists in the nucleon excitation spectrum. The Roper resonance, with quantum numbers $J^P=1/2^+$, is generally identified as the first radial excitation of the nucleon, whereas $N^*(1535)$, with $J^P=1/2^-$, is commonly interpreted as its first orbital excitation. In conventional quark models, the positive-parity radial excitation is expected to lie above the negative-parity orbital excitation in mass. Experimentally, however, the opposite ordering is observed. This discrepancy suggests that the physical resonance masses cannot be identified directly with the masses of uncoupled three-quark cores, and that dressing effects arising from the surrounding meson--baryon cloud may play a non-negligible role. HEFT therefore provides an important theoretical framework for investigating the structure of excited nucleon states.

The first nucleon excitation investigated within HEFT was $N^*(1535)$~\cite{Liu:2015ktc}. In this description, the resonance was assumed to contain a bare three-quark core coupled to the $\pi N$ and $\eta N$ channels, while direct interactions between the two meson--baryon channels were also included. The resulting infinite-volume resonance pole was consistent with the PDG determination available at the time. The Hamiltonian constrained by the infinite-volume experimental data was then formulated in finite volume, allowing the study to successfully predict several sets of lattice QCD energy levels for spatial extents of approximately $2~\mathrm{fm}$ and $3~\mathrm{fm}$. Furthermore, an analysis of the eigenstate composition indicated that, near the physical pion mass, the eigenstate associated with $N^*(1535)$ has a bare-state probability of approximately 50\%, with the remaining contribution distributed between the $\pi N$ and $\eta N$ channels.

More recently, lattice QCD calculations of the magnetic moments of negative-parity nucleon excitations found that two lattice states associated with $N^*(1535)$ and $N^*(1650)$ exhibit magnetic moments close to the predictions of the constituent-quark model. Since two bare three-quark cores carrying the same quantum numbers can mix through their couplings to common meson-baryon channels, Abell et al. constructed a two-bare-state Hamiltonian within HEFT~\cite{Abell:2023nex}. In this framework, both $N^*(1535)$ and $N^*(1650)$ were assumed to contain three-quark cores coupled to the $\pi N$, $\eta N$, and $K\Lambda$ channels. In infinite volume, the model produced two resonance poles in good agreement with the corresponding PDG determinations. In finite volume, the authors first used lattice QCD results at a spatial extent of approximately $3~\mathrm{fm}$ to constrain the slopes governing the chiral extrapolation of the two bare-state masses. One of the extracted slopes was found to be very close to that obtained in the earlier single-bare-state analysis of $N^*(1535)$. After all model parameters had been fixed, the predicted energy levels at spatial extents of approximately $2~\mathrm{fm}$ and $4.05~\mathrm{fm}$ were also found to be consistent with the corresponding lattice QCD spectra.

In contrast to the case of $N^*(1535)$, lattice QCD calculations employing three-quark interpolating fields found a nucleon radial excitation near $1.8~\mathrm{GeV}$, well above the physical Roper mass of approximately $1.44~\mathrm{GeV}$~\cite{Mahbub:2013ala, Alexandrou:2014mka}. To investigate this discrepancy, an early study of the $\pi N$, $\pi\Delta$, and $\sigma N$ coupled channels compared three scenarios: a bare Roper basis state, no bare baryon basis state, and a bare nucleon basis state~\cite{Liu:2016uzk}. All described the $\pi N$ phase shifts and inelasticities, but their finite-volume spectra and eigenvector compositions differed. The bare-Roper scenario predicted low-lying eigenstates with significant bare-state components that were not observed in the lattice calculations considered. The analysis favored the bare-nucleon scenario for the larger-volume lattice results. At light pion masses, the relevant excited eigenstates were dominated by meson-baryon components, with small bare-nucleon contributions. These contributions helped explain the excitation of particular states by three-quark interpolating fields.

In a subsequent HEFT study, Wu et al. coupled a bare baryon basis state, representing a quark-model-like radial excitation, to the same three channels~\cite{Wu:2017qve}. Two parameter sets described the $\pi N$ scattering data but gave different finite-volume spectra. Comparison with lattice results obtained employing three-quark and meson-baryon interpolating fields favored the scenario with a bare-state mass of approximately $2~\mathrm{GeV}$. In this scenario, the Roper pole was generated predominantly by meson-baryon rescattering, whereas finite-volume eigenstates with large bare-state components appeared at higher energies. A lower bare-state mass produced low-lying eigenstates with large bare-state components that were inconsistent with the lattice results considered.

\subsection{$\Delta(1232)$, $\Delta(1600)$ and $\Delta(1920)$}
The $\Delta(1232)$, $\Delta(1600)$ and $\Delta(1920)$ are identified through $P_{33}~\pi N$ partial-wave and coupled-channel analyses, although the higher and broader states remain less precisely determined \cite{ParticleDataGroup:2026aaa,Arndt:2006bf}. A puzzle analogous to the Roper resonance also appears in the $\Delta$ spectrum. The $\Delta(1600)$, traditionally identified as the first radial excitation of the $\Delta(1232)$, lies below the negative-parity $\Delta(1700)$, and substantially below the predominantly three-quark $2S$ state determined in lattice QCD, which is located near $2.15~\mathrm{GeV}$. This discrepancy suggests that the physical origin of the $\Delta(1600)$ may differ from that of a conventional radial excitation. Using a model containing only a single bare state, Hockley et al. obtained two infinite-volume resonance poles associated with the $\Delta(1232)$ and $\Delta(1600)$, respectively~\cite{Hockley:2024ipz}. This result indicates that the emergence of the $\Delta(1600)$ does not require the introduction of an additional radially excited bare state. Upon including a second bare state, the full model generated a third pole associated with the $\Delta(1920)$.

The finite-volume eigenstate-composition analysis further showed that the first and second bare states dominate in the energy regions associated with the $\Delta(1232)$ and $\Delta(1920)$, in good agreement with the energy spectrum reported by the CSSM Collaboration~\cite{Hockley:2023yzn}. By contrast, the eigenstates in the $\Delta(1600)$ region are characterized predominantly by strong mixing between the $\pi N$ and $\pi\Delta$ components, with only a small bare-state contribution. On this basis, the authors favor an interpretation of the $\Delta(1600)$ as a predominantly dynamically generated resonance. This picture closely resembles the HEFT interpretation of the Roper resonance.

\subsection{$\Lambda(1405)$ and $\Lambda(1670)$}
The $\Lambda(1405)$ and $\Lambda(1670)$ are constrained primarily through low-energy $K^-p$ reactions and coupled $\pi\Sigma\text{--}\bar{K} N$ analyses, supplemented by recent measurements of kaonic hydrogen, correlation functions, and $\pi \Sigma$ spectra~\cite{ParticleDataGroup:2026aaa, SIDDHARTA:2011dsy,ALICE:2021szj,J-PARCE31:2022plu}. 
The $\Lambda(1405)$ is the lightest well-established negative-parity baryon, lying below the corresponding non-strange nucleon excitations in a manner that is difficult to reconcile with a simple quark-model ordering. This suggests that it may not be a conventional three-quark excitation, but instead a dynamically generated molecular state. This possibility was investigated by comparing HEFT models with and without a bare baryon basis state, using the coupled $\pi\Sigma$, $\bar{K}N$, $\eta\Lambda$, and $K\Xi$ channels~\cite{Liu:2016wxq}. Both scenarios successfully described the $K^-p$ scattering cross sections and produced the characteristic two-pole structure of the $\Lambda(1405)$, with nearly identical results. These similar scattering fits alone do not distinguish the two scenarios.

\begin{figure}[H]
\subfloat[\centering]{\includegraphics[width=6.5cm]{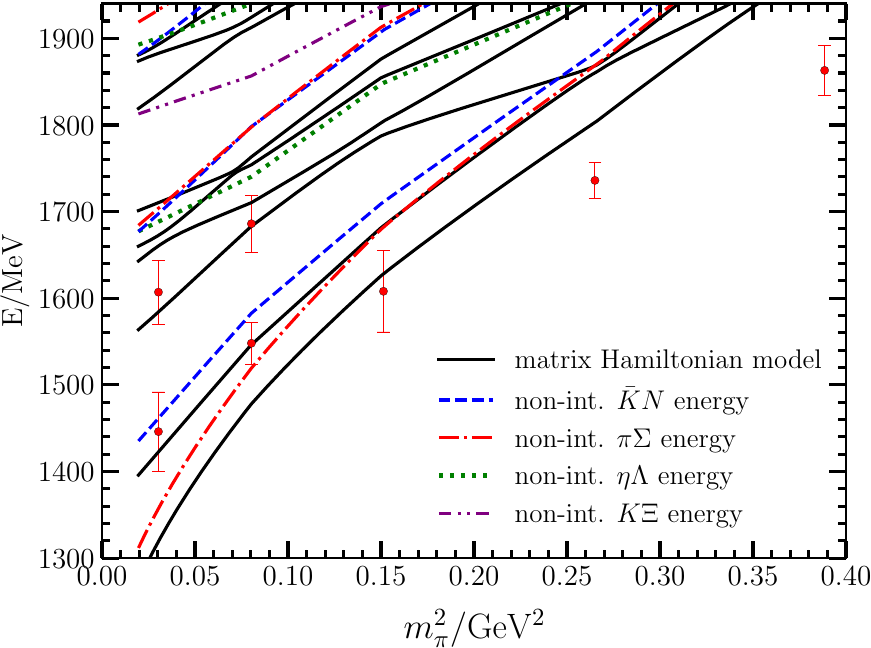}}
\subfloat[\centering]{\includegraphics[width=6.5cm]{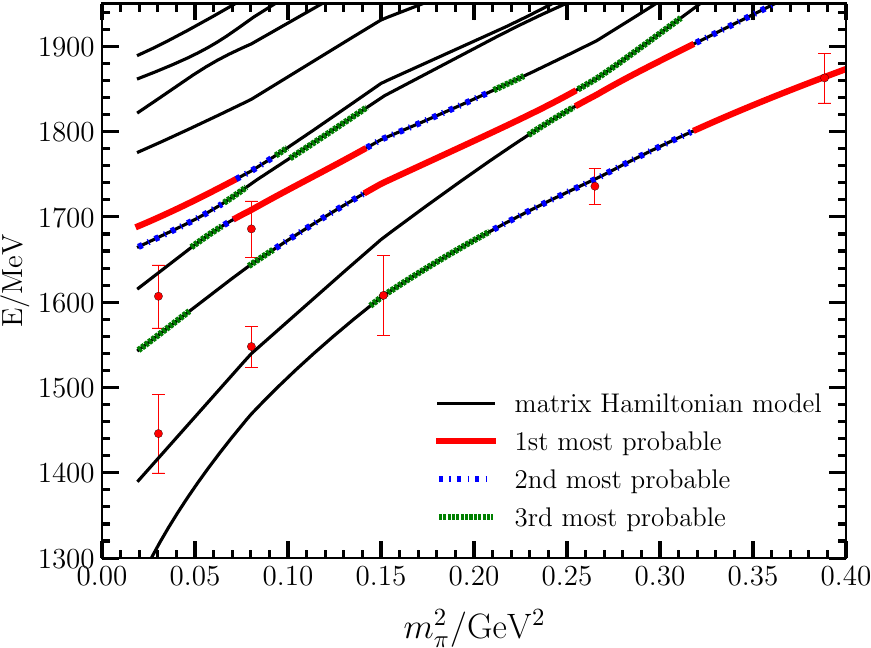}}
\caption{Finite-volume energy levels in the $J^P=1/2^-$ $\Lambda$ sector as functions of $m_\pi^2$, for a spatial extent of $L\simeq3~\mathrm{fm}$. The solid curves represent the finite-volume HEFT eigenenergies. Panel (a) shows the scenario without a bare state. The dashed curves in different colors represent the noninteracting $\bar{K}N$, $\pi\Sigma$, $\eta\Lambda$ and $K\Xi$ energy levels. Panel (b) shows the scenario containing a bare state. The red solid, blue dot-dashed, and green dotted curves indicate the states with the largest, second-largest, and third-largest bare-state contributions, respectively. The LQCD data points denote results from the CSSM Collaboration~\cite{Menadue:2011pd,Hall:2014uca} in $2+1$ flavor QCD~\cite{PACS-CS:2008bkb}. The figure is taken from Ref.~\cite{Liu:2023xvy}.\label{fig2}}
\end{figure}

The two scenarios differ more clearly in their pion-mass dependence. As shown in Fig.~\ref{fig2}(a) and (b), at larger $m_\pi$ the model without a bare state can no longer naturally reproduce the low-lying state observed in lattice QCD, whereas the model containing a bare state continues to provide a consistent description. In the model with a bare state, the heavy-pion eigenstate is about 80\%--90\% bare. As $m_\pi$ decreases, avoided level crossings redistribute the components among eigenstates. Near the physical point, the second eigenstate at $L\simeq2.9~\mathrm{fm}$ is about 90\% $\bar{K}N$. The study therefore favors an interpretation of the $\Lambda(1405)$ as predominantly a $\bar{K}N$ molecular state. Meanwhile, the conventional orbitally excited $uds$ three-quark configuration is not absent, but instead evolves into higher-lying $\Lambda$ excitations as the quark masses decrease.

The $\Lambda(1670)$ may represent the higher-lying state into which the conventional $uds$ three-quark configuration evolves. This possibility was tested by comparing HEFT models with and without a bare baryon basis state against scattering data over an extended momentum range~\cite{Liu:2023xvy}. Both scenarios simultaneously described the $\Lambda(1405)$ and $\Lambda(1670)$, producing similar cross-section line shapes and pole structures. Their finite-volume spectra, however, differed markedly. As in the case of the $\Lambda(1405)$, the model without a bare state reproduced the low-energy scattering states only near small $m_\pi$, while deviating substantially from the lattice QCD results at larger $m_\pi$. After introducing a bare state, the same Hamiltonian provided a good description of the principal energy levels reported by the CSSM~\cite{Menadue:2011pd,Hall:2014uca} in $2+1$ flavor QCD~\cite{PACS-CS:2008bkb} and BaSc~\cite{BaryonScatteringBaSc:2023ori,BaryonScatteringBaSc:2023zvt} collaborations across a broad range of quark masses. The eigenstate-composition analysis further indicated that the $\Lambda(1670)$ contains a significant three-quark core strongly dressed by the relevant meson--baryon channels, in clear contrast to the predominantly molecular structure inferred for the $\Lambda(1405)$.

\subsection{ Low-Lying $\Sigma$ Resonances}
Experimental information remains limited in the $I=1$, $J^P=1/2^-$ $\Sigma$ sector and is derived primarily from $\bar{K}N$ coupled-channel analyses and selected hadronic and heavy-hadron decay processes~\cite{Kamano:2015hxa,Wu:2009tu,Wang:2024jyk}. It is unclear whether a low-mass structure exists near $1.4~\mathrm{GeV}$ and whether the $1.6\text{--}1.75~\mathrm{GeV}$ region contains one or two resonances. Finite-volume lattice QCD spectra therefore provide an important means of discriminating among different interpretations. Ma et al. considered five $I=1$ $S$-wave channels, namely $\pi\Lambda$, $\pi\Sigma$, $\bar{K}N$, $\eta\Sigma$, and $K\Xi$, and compared a purely dynamically generated meson--baryon scenario with one that additionally contains a three-quark bare state~\cite{Ma:2025wbz}.

The infinite-volume analysis exhibited a limitation similar to that encountered in the preceding studies: both scenarios provided a broadly acceptable description of the available cross sections and therefore could not by themselves establish whether a three-quark core is required. Nevertheless, the pole structures differed, with one pole obtained in the model without a bare state and two poles in the model containing one. A pronounced cusp was also found near the $\bar{K}N$ threshold in the scattering amplitudes.

In finite volume, comparison with lattice QCD spectra from the BGR Collaboration~\cite{Engel:2013ig} showed that the results at smaller $m_\pi$ favor the scenario containing a bare state. At larger $m_\pi$, however, the uncertainties and the limited number of available energy levels prevent a clear distinction between the two models. 

\subsection{$\Omega(2012)$}
The $\Omega(2012)^{-}$ was first observed by Belle in the $\Xi^{0}K^{-}$ and $\Xi^{-}K_{S}^{0}$ channels \cite{Belle:2018mqs}, with subsequent measurements constraining its couplings to $\Xi\bar{K}$ and $\Xi(1530)\bar{K}$ \cite{Belle:2019zco,Belle:2022mrg}. Further evidence has been reported by BESIII and ALICE \cite{BESIII:2024eqk,ALICE:2025atb}. Its proximity to the $\Xi(1530)\bar{K}$ threshold, only about $16.5~\mathrm{MeV}$ above the resonance mass, suggests an important role for this near-threshold channel in the $\Omega(2012)^{-}$ structure.

Experimental information on coupled-channel interactions in the $\Omega$ sector is limited, since $K^-p$ scattering cannot produce an $S=-3$ $\Omega^-$ resonance in the $s$ channel. In particular, the lack of precise $\bar{K}\Xi$ scattering data prevents a direct determination of the $\Omega_{0}\text{--}\Xi\bar{K}$ and $\Omega_{0}\text{--}\Xi(1530)\bar{K}$ transition vertices.

Han et al. therefore investigated the negative-parity $\Omega$ spectrum within Hamiltonian effective field theory by coupling a bare three-quark $\Omega$ core to the $\Xi\bar{K}$ and $\Xi(1530)\bar{K}$ channels \cite{Han:2025gkp}. The transition vertices were constrained using the ${}^{3}P_{0}$ model \cite{Wang:2018hmi}, following a strategy analogous to that previously employed in coupled-channel studies of the near-threshold $D_s$ states and the $P$-wave $B_s$ spectrum \cite{Yang:2021tvc,Yang:2022vdb}. The vertex function coupling the bare $\Omega$ core to the two channels $\alpha$ is given by
\begin{equation}\label{eq:gfin}
G^{J}_{\alpha,\Omega_{0}}(k)= \mathcal{M}^{JL}_{\alpha}(k) \exp\left(-\frac{k^{2}}{2\Lambda^{2}}\right),
\end{equation}
where $\mathcal{M}^{JL}_{\alpha}(k)$ is the corresponding partial-wave amplitude obtained from the ${}^{3}P_{0}$ model, and the exponential regulator suppresses high-momentum contributions to the vertices.

The remaining Hamiltonian parameters, including the bare-state mass and its pion mass dependence, were constrained using the negative-parity $\Omega$ spectrum calculated by the CSSM Collaboration with smeared three-quark interpolating operators on $N_f=2+1$ PACS-CS gauge configurations at a spatial extent of $L\simeq 2.9~\mathrm{fm}$ \cite{Hockley:2024aym,PACS-CS:2008bkb}.

\begin{figure}[H]
\center
\includegraphics[width=3.2in]{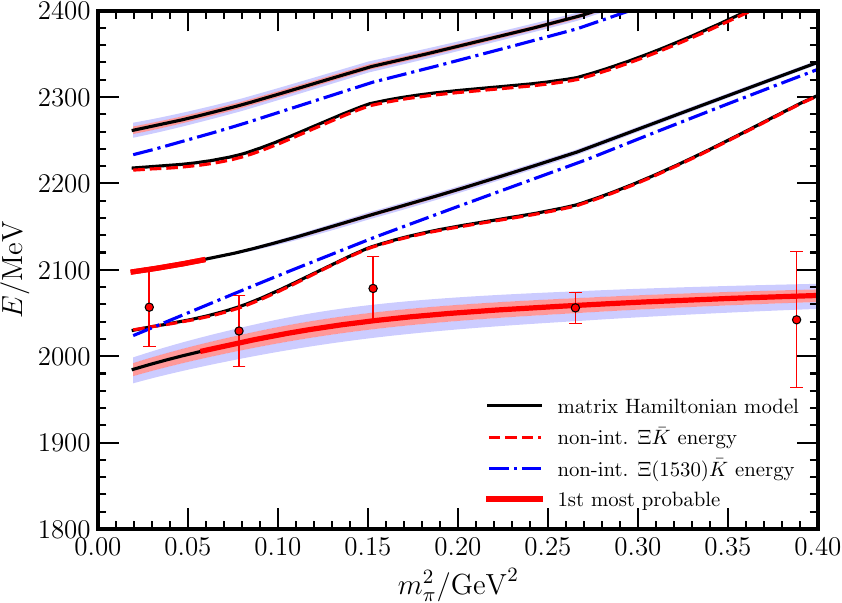}
\caption{Finite-volume energy spectrum of the $J^{P}=3/2^{-}$ system as a function of $m_\pi^2$. The red solid line denotes the eigenstate with the largest bare component; the broken curves show the noninteracting $\Xi\bar{K}$ and $\Xi(1530)\bar{K}$ energies. Brown and blue bands indicate $10\%$ and $20\%$ variations of the ${}^3P_{0}$ coupling $\gamma$. The lattice data are from the CSSM Collaboration~\cite{Hockley:2024aym} and were obtained in QCD with 2+1 dynamical quark flavors~\cite{PACS-CS:2008bkb}. The figure is taken from Ref.~\cite{Han:2025gkp}.}
\label{ben3}
\end{figure}

HEFT thus combines quark-model input with lattice QCD spectra and experimental constraints within a unified finite- and infinite-volume Hamiltonian framework. Figure~\ref{ben3} shows the finite-volume spectrum of the $J^{P}=3/2^{-}$ $\Omega$ system as a function of $m_{\pi}^{2}$. The thick red curves denote eigenstates with the largest bare-$\Omega$ components, which are expected to have enhanced overlap with the local three-quark interpolating operators used in the lattice calculation. All five lattice QCD levels are consistent, within uncertainties, with the corresponding HEFT eigenstates.
In the two-body approximation with a stable $\Xi(1530)$, the $J^{P}=3/2^{-}$ infinite-volume $T$ matrix exhibits a pole at $(2012\pm 8)-i(1.6\pm0.2)~\mathrm{MeV}$ corresponding to a width of $\Gamma=3.2\pm0.4~\mathrm{MeV}$ and supporting its identification with the observed $\Omega(2012)^{-}$. In contrast, the $J^{P}=1/2^{-}$ solution is substantially broader and may instead be associated with the $\Omega(2109)^{-}$ reported by BESIII. For the favored $J^{P}=3/2^{-}$ assignment, the near-threshold $S$-wave $\Xi(1530)\bar{K}$ channel provides an essential contribution to the real part of the pole, whereas the open $D$-wave $\Xi\bar{K}$ channel predominantly generates its imaginary part and hence the decay width. These results favor a $J^{P}=3/2^{-}$ assignment and characterize the $\Omega(2012)^{-}$ as a coupled system containing a three-quark core and meson-baryon components rather than a purely undressed three-quark state or a purely molecular configuration.

\section{Complex Systems and Theoretical Extensions}
\label{secc}

\subsection{Moving Frames and Elongated Volumes}
Most of the HEFT studies discussed above were formulated in a cubic finite volume in the rest frame, where the two-particle relative momenta are discretized in units of $2\pi/L$. Increasing $L$ is a direct way to obtain a finer momentum resolution and improve precision, but it also substantially increases the computational cost. An alternative is to employ elongated volumes and moving frames with nonzero total momentum, which likewise provide a denser sampling of the scattering amplitude. The price, however, is a reduction of the spatial symmetry, leading to more complicated partial-wave mixing and requiring a consistent momentum mapping between the interactions defined in the center-of-mass frame and the boundary conditions imposed in the moving frame.

Following early extensions of Lüscher's method to asymmetric boxes~\cite{Li:2003jn,Feng:2004ua}, later studies developed scattering formalisms for elongated boxes~\cite{Lee:2017igf,Li:2018tzx} and employed these geometries in lattice QCD~\cite{Guo:2016zos,CLQCD:2007rcz,CLQCD:2009nvn,Pelissier:2012pi,Guo:2018zss,Culver:2019qtx,Culver:2019vvu}. Li et al. extended HEFT to elongated and moving volumes, including configurations with the elongation direction aligned with the total momentum~\cite{Li:2021mob}. Denoting the elongation direction by $\boldsymbol{d}_\eta$ and the elongation factor by $\eta$, the allowed finite-volume momenta can be written as
\begin{eqnarray}
\boldsymbol{k}=\frac{2\pi}{L}( \boldsymbol{n}_{\bot}+\frac{1}{\eta}\boldsymbol{n}_{\parallel}).
\end{eqnarray}
For $\eta>1$, the momentum spacing along the elongation direction is reduced. Suppose further that the system carries a total momentum
\begin{eqnarray}
\boldsymbol{P}=\frac{2\pi}{L}(\boldsymbol{d}_\perp+\frac{\boldsymbol{d}_\parallel}{\eta}),\quad\boldsymbol{d}\in\mathbb{Z}^3.
\end{eqnarray}

For a fixed spatial volume, different choices of $\boldsymbol{d}$ lead to different allowed center-of-mass energies, thereby substantially increasing the number of sampling points for the scattering phase shifts. When $\boldsymbol{d}_\eta\parallel\boldsymbol{P}$, the symmetry reductions induced by the elongation and the moving frame remain mutually compatible. Li et al. adopted an energy-independent momentum-mapping prescription, removing the explicit energy dependence that would otherwise arise in the moving-frame Hamiltonian. By including the associated Jacobian factor, interactions parametrized in the center-of-mass frame could be incorporated naturally to generate finite-volume spectra in moving frames.

To accommodate the more complicated partial-wave mixing, the authors also developed an efficient method for organizing the Hilbert space and constructing a symmetry-adapted orthonormal basis of minimal dimension. The framework achieved initial success in an application to $I=2$ $\pi\pi$ scattering. When spectra from the rest frame and several moving frames were fitted using a common Hamiltonian, the extracted central values were mutually consistent, while the inclusion of moving-frame data significantly reduced the uncertainties in the phase shifts. In subsequent work, Li et al. derived the formulation from the general moving-frame quantization condition and systematized the earlier momentum mapping into the LWLY transformation \cite{Li:2024zld}. They further demonstrated that the LWLY, RG, and KSS prescriptions are equivalent with respect to power-law finite-volume effects. These developments substantially broadened the applicability of HEFT to modern lattice QCD spectra.

\subsection{The N-Body Formalism}

Many of the HEFT studies discussed above include quasi-two-body channels such as $\pi\Delta$ and $\sigma N$. Since the $\Delta$ and $\sigma$ are themselves unstable, however, the underlying physical channel is actually the three-body $\pi\pi N$ channel. Without treating the three-body states explicitly, their more complicated analytic structure may not be fully captured. To address this issue, Yu et al. developed a more general nonperturbative Hamiltonian framework (NPHF) for multihadron systems~\cite{Yu:2026qlt}.

For a conventional two-body HEFT calculation at zero total momentum, only one independent relative momentum is required. A three-body system, by contrast, requires at least two independent relative momenta, causing the dimension of the Hamiltonian matrix to grow rapidly. For example, with the momentum cutoff $\lVert\boldsymbol{n}_i\rVert^2\leq 20$, the number of two-body momentum configurations is approximately 400, whereas the corresponding number for a three-body system increases to about 70,000. Direct diagonalization of the resulting Hamiltonian matrix therefore becomes computationally challenging.

Within NPHF, a single-particle bare state is denoted by $|B,\sigma\rangle$, while a general $N$-particle state is written as $|\alpha^{(N)}_a(\boldsymbol{k}_1,\sigma_1;\ldots;\boldsymbol{k}_N,\sigma_N)\rangle$, where the subscript $a$ labels a column of an irreducible representation of the permutation group. The finite-volume Fourier factor arising from periodic boundary conditions is then generalized to a particle-number-dependent form~\cite{Yu:2026qlt}:
\begin{eqnarray}
&&\frac{\langle B|V_F|\alpha_L^{(M)}\rangle}{\langle B|V|\alpha^{(M)}\rangle}=(\frac{2\pi}{L})^\frac{3(M-1)}{2},\\
&&\frac{\langle\beta_L^{(M)}|V_F|\alpha_L^{(N)}\rangle}{\langle\beta^{(M)}|V|\alpha^{(N)}\rangle}=(\frac{2\pi}{L})^\frac{3(M+N-2)}{2}.
\end{eqnarray}

NPHF decomposed the Hilbert space according to lattice irreducible representations and momentum orbits. Each subspace was projected onto a definite lattice irrep, after which an overlap matrix was constructed to identify and remove linear dependencies among the projected states, yielding sets of orthonormal basis states. With the simplifications provided by the overlap-matrix analysis and the Wigner--Eckart theorem, the total dimension of these symmetry-adapted bases is substantially smaller than the number of all possible momentum configurations.

Yu et al. presented an initial application of NPHF to the dynamics of the $\omega$ meson. The physical $\omega$ decays predominantly into three pions~\cite{Yan:2024gwp} through an off-shell $\rho\pi$ configuration, with $\rho\to\pi\pi$ occurring as a two-body subprocess within the overall $\omega\to3\pi$ decay. The calculation must therefore treat simultaneously the $I=0$ transition chain $\omega_0\longleftrightarrow\rho\pi\longleftrightarrow\pi\pi\pi$ and the $I=1$ process $\rho_0\longleftrightarrow\pi\pi$. Since the $\rho$ dynamics are shared by the two sectors, the two-pion and three-pion lattice spectra must be fitted simultaneously using a common Hamiltonian. In the study by Yu et al., a single Hamiltonian with shared couplings and pion-mass-dependent bare masses successfully reproduced the finite-volume spectra in both isospin sectors at two pion masses and across multiple spatial volumes.

\subsection{Pion Photo- and Electroproduction of Nucleon Excited States}

Pion photoproduction provides information on nucleon excitations that is complementary to that obtained from hadronic scattering. In particular, the electromagnetic couplings can help distinguish states that couple only weakly to the $\pi N$ channel. HEFT was extended to the reaction $\gamma N\to\pi N$ by separating the production amplitude into an electromagnetic transition and the subsequent strong final-state interactions~\cite{Zhuge:2024iuw}. The former was constructed from effective Lagrangians, while the latter employed the half-off-shell meson--baryon amplitudes obtained in earlier HEFT analyses. Thus, the parameters governing the strong interactions remained constrained by the $\pi N$ scattering data and finite-volume lattice spectra.

For the $S_{11}$ partial wave, the calculation included the $\pi N$, $\eta N$, and $K\Lambda$ channels together with two bare states associated with $N^*(1535)$ and $N^*(1650)$. The inclusion of the second bare state improved the electric multipole amplitude $E_{0+}$ above approximately $1.5~\mathrm{GeV}$, and the bare-state terms were found to be essential, in accord with the substantial three-quark components inferred for these two resonances. The magnetic multipole amplitude $M_{1-}$ in the $P_{11}$ wave was studied using the $\pi N$, $\sigma N$, and $\pi\Delta$ channels. Here the quark-model-like bare state lies near $2~\mathrm{GeV}$ and makes only a small contribution in the Roper region. The resulting amplitude is comparable to those obtained in other dynamical coupled-channel analyses and supports the HEFT picture of $N^*(1440)$ as a resonance generated predominantly by meson--baryon rescattering.

More recently, the same framework was extended to near-threshold pion electroproduction and confronted directly with lattice QCD multipole amplitudes at nearly physical $m_\pi$~\cite{Zhuge:2026dxl}. The virtual-photon momentum dependence was incorporated through the nucleon and pion electromagnetic form factors, while the strong final-state interactions included the $\pi N$, $\eta N$, and $K\Lambda$ channels. Using the photoproduction parameters and empirical $\pi NN$ input, the calculated real part of $E_{0+}$ agreed with the lattice results within their quoted uncertainties. Near threshold, more than 85\% of the strength arises from the contact term. The closed $\eta N$ and $K\Lambda$ channels contribute only a few percent, but remain relevant for a precision determination of the amplitude.

This study also clarified how finite-volume electroproduction amplitudes may be related to physical observables. Under a separable-potential assumption, the authors derived a complex finite-volume correction factor for the first excited state considered in their analysis, allowing both the real and imaginary parts of the infinite-volume amplitude to be recovered, while the usual Lellouch--L\"uscher factor determines only its magnitude~\cite{Lellouch:2000pv}. At the lowest level, however, the finite-volume energy is shifted below the physical $\pi N$ threshold, and the additional energy extrapolation depends on both the electromagnetic transition and the strong interactions. A dynamical treatment of the energy dependence is therefore required for a controlled threshold correction. Since the finite-volume effects were found to decrease for the first excited level, future lattice calculations at higher energies should provide stronger constraints on the electromagnetic structure of nucleon resonances.

\section{Summary}
\label{secd}

Hamiltonian effective field theory provides a unified dynamical framework for relating infinite-volume scattering amplitudes to finite-volume lattice QCD spectra by incorporating bare single-hadron configurations and relevant multihadron channels within a common Hilbert space. Once constrained by experimental scattering data or lattice spectra, the same Hamiltonian can describe resonance poles in infinite volume and discrete energy levels in finite volume, while its eigenvectors provide model-dependent insight into the distribution of bare-state and multihadron components among lattice eigenstates. The applications reviewed here demonstrate that scattering observables alone are often insufficient to determine resonance structure uniquely, as different Hamiltonian scenarios may reproduce similar phase shifts and pole positions while predicting distinct finite-volume spectra. Lattice QCD therefore provides crucial complementary constraints, as illustrated by studies of the Roper resonance, the $\Lambda(1405)$, and the $\Lambda(1670)$. Eigenvector analyses further help explain why resonances dominated by meson-baryon dynamics, such as the Roper resonance, the $\Delta(1600)$, and the $\Lambda(1405)$, can have weak overlap with conventional three-quark interpolating operators, whereas predominantly quark-model-like excitations may appear at considerably higher energies.

Recent developments have substantially broadened the scope of Hamiltonian methods. Moving frames and elongated volumes provide additional finite-volume energy levels and denser sampling of scattering amplitudes, while general $N$-body formulations enable the explicit treatment of three-body dynamics. Extensions to electromagnetic production processes offer further, complementary probes of resonance structure. Overall, HEFT has evolved from a phenomenological tool for specific coupled-channel systems into a versatile framework connecting experimental information with lattice QCD. With continuing advances in both experiment and lattice simulations, it provides a promising approach for clarifying the dynamical origins of hadron resonances, testing low-energy QCD dynamics, and improving our understanding of the baryon spectrum and more complex hadronic systems.

\funding{This work is supported by the National Natural Science Foundation of China under Grants No. 12335001, No. 12175091, and No. 12247101, the “111 Center” under Grant No. B20063, and the innovation project for young science and technology talents of Lanzhou city under Grant No. 2023-QN-107. }

\acknowledgments{The authors would like to express their sincere appreciation to all colleagues whose contributions have informed and supported the work presented in this article.}

\conflictsofinterest{The authors declare no conflicts of interest.}

\reftitle{References}

\bibliography{reference}%

\end{document}